 \documentclass[mathleft]{an}
\usepackage{graphicx}
\usepackage{times}
\begin{document}

\title{The Effective Tidal Viscosity in Close Solar-Type Binaries}
 
     \author{Itzhak Goldman }
 \institute{Department of Astronomy and Astrophysics, Tel Aviv University, Tel Aviv, 69978, Israel and Afeka,Tel Aviv Academic College of Engineering,
218 Bnei Efraim, Tel Aviv 61533, Israel, \email{goldman@wise.tau.ac.il}}

\keywords{binaries: close --- stars: solar-type --- convection --- turbulence}

\abstract{
A major problem confronting the understanding of tidal evolution of close solar-type binaries is the inefficiency of the turbulent  convection. The  value  of the effective viscosity estimated, in the framework of the mixing length theory (MLT), implies  circularization  timescales   which are almost  two orders of magnitude longer than observed. Moreover,     the reduction of
 the effective viscosity due to the fast time-variation of the  tidal shear   in short period binaries,  increases the discrepancy to about  three orders of magnitude. This state of affairs has motivated suggestions that tidal orbital evolution, notably circularization occurs mainly during the pre-main-sequence phase. However, observational data accumulated over the recent decades imply that circularization does occur during the the main-sequence phase (Mazeh 2008).
\newline 
In this work, we  examine the possibility that the apparent inefficiency of turbulent convection   is merely a shortcoming of  MLT  approach.  Indeed,  a recent 3D numerical simulation (Penev et al. 2007),  suggests that the true convective viscosity is probably larger than the MLT value and that the reduction due to the time-variation of the shear is not drastic. 
We  employ a  model for stellar turbulent convection (Canuto, Goldman \&  Mazzitelli   1996) to evaluate the effective viscosity both for a steady for  and time dependent tidal shear. 
\newline 
 The model is physically based, self-consistent, and accounts for the full spectrum of the turbulent eddies. It has been found   advantageous, compared to the MLT, in many applications.
We use an  analytic approximation to  the turbulent spectrum to obtain the reduction of the efficiency due to the time-variation of the tide. The results are:\newline (i) an enhanced effective viscosity ( by a factor of $\sim 4.5$) and more importantly\newline (ii) only a mild  reduction due to the time-variation of the tidal shear. \newline Overall,  for binaries with orbital period of  $15$ days the discrepancy is
 "only" a factor of $\sim 30$  down from a factor of $\sim 1000$. These encouraging results   should motivate  an investigation  of  rigorous non-analytic solutions.}
 
\maketitle

\section{Introduction}

About two decades ago   there has been a revival of interest 
  in the  tidal  orbital
     evolution of close binary systems. This was mainly due to the
     availability of large samples of spectroscopic binaries, made
     possible by the (then) new generation of stellar speedometers in full
     operation at several observatories (Latham et al. 1988; Mathieu and Mazeh 1988);
  Latham et al. 1992a; Latham et al. 1992b;  Mathieu 1992; Mathieu et al. 1992).
More recently, the  interest in the subject has been renewed by Latham et al. (2002), Mathieu et al. (2003), Mathieu (2005), Mazeh  (2008), Meibom (2005), \newline Meibom and Mathieu  (2004, 2005), 
Meibom et al. (2006), Meibom et al.  (2007). 

 From the theoretical point of view, the central issue is the nature of the mechanism responsible for the
dissipation of the energy in the tide and for the transfer of angular momentum between the binary orbit and the stellar rotations.  A successful model should predict the dependence of observed parameters   the binary orbital parameters. 

For these binaries,  it is commonly believed that the   dominant
physical process controlling energy dissipation is  the turbulent convection   in the stellar envelop. In this equilibrium tide model  (Alexander 1973; Lecar et. al., 1976; Zahn 1977; Hut 1981, 1982; Eggleton et al. 1998), the tidal force induces a sheared velocity field in each of the stars.  
 The  turbulent convection interacts with the shear drawing from it energy. Technically, the
effect of the turbulence is represented by an enhanced   effective diffusivity-- the turbulent viscosity which is orders of magnitude  larger than the microscopic viscosity. 
 Within this framework, the
    circularization timescale for late-type stars with convective
    envelopes is proportional to the orbital period of the binary to the
    power 16/3  (Zahn 1977). 
 
  Zahn  (1989) called attention to the fact that   the tidal effective viscosity is reduced because the orbital period is shorter than the  
typical timescale of the turbulent convection largest eddies. Thus, the effect of the large eddies is   averaged out and  only smaller   eddies, with shorter timescales can contribute to the viscosity. This implies a reduction of the viscosity and thus implies an increase of the circularization timescales. According to Zahn  (1989) the reduction is linear in the period thus implying that the circularization timescale  is proportional to the orbital period  to the
    power 13/3. As a result, Zahn and Bouchet  (1989)   argued that  most of the orbital
    circularization should occur during the pre-main-sequence phase. 

Goldman \& Mazeh  (1991) concluded,  
  in analogy with kinetic microscopic viscosity,  that the reduction should be   quadratic in the period implying 
  that the circularization \newline
 timescale 
    is proportional to the orbital period   to the
    power 10/3. Such a quadratic reduction has been proposed by 
Goldreich \& Keeley  (1977)  
  in the context of the damping of solar pulsations by
  turbulence   convection. The observations available at the time    favored an index of 10/3. This led Goldman\& Mazeh  (1992) to suggest that the model is essentially correct and  the excessively  long timescales   reflect the MLT shortcomings, notably not representing the true topology of the turbulent convection and not accounting for the entire turbulent spatial energy spectrum.  Moreover, since the MLT is basically a descriptive framework, the
expressions used for the turbulent viscosity and for the turbulent timescales are based on 
dimensional analysis and the  numerical values of the various parameters are thus uncertain.

The reduction of the turbulent convective viscosity has been readdressed  by
Goodman and Oh  (1997). They obtained that  the reduction increases gradually  as 
 the orbital period gets shorter; it is quadratic in the limit of very short orbital periods.
Using the MLT, they found that the ensuing circularization timescales for an orbital period  in the range of 15 - 10 days are longer by an additional factor of 10. Overall, for a 15 day binary period, the predicted timescale  is too long  by a factor close to  $1000$.

Recent observations  (Meibom \& Mathieu   2004; 2005;  
Meibom et al. 2006;    Meibom et al. 2007; Mazeh 2008) reveal a more complex picture than that of the past.  Tidal evolution does take  place during the main sequence phase. However,  the functional dependence of the tidal timescale is not
clear; the  observational uncertainties do not allow a
  determination of the index of the power-law, if indeed  there is a description by a single power-law. 

 An interesting paper   by  Penev et al. ( 2007) reports the results of a 3D numerical simulation  suggest  that the true convective viscosity is probably larger than that indicated by the MLT for the relevant orbital periods.

\section{Motivation for Revisiting the Problem}

The new observational data mentioned in the last paragraph, 
seem to imply that circularization {\it does occur} during the main sequence phase. Since, at least qualitatively, turbulent convection is the natural candidate for supplying the effective viscosity, we wish
to reconsider the problem in a  theoretical framework more consistent than the MLT. Support
in this direction is given by Penev et al. ( 2007).

We use a model for  turbulent convection of Canuto, \newline  Goldman, \&  Mazzitelli (1996)  (CGM96) which  is based on previous  turbulence  models
  (Canuto et al 1984; Canuto and Goldman 1985; Canuto et al. 1987;   Canuto et al. 1988).
The CGM96 model  has been   applied to a large body (over 130 citations in ADS)  of    astrophysical problems, spanning a variety of aspects of stellar convection: stellar structure, stellar evolution, age of globular clusters, helio seismology, and blue-edge of   DBV pulsating white dwarfs .

The CGM96 model is a 
self-consistent model for turbulent convection which accounts for
the {\it full spectrum} of the turbulent convective eddies. It constitutes  
a
significant improvement over the mixing length theory  (MLT) which
is a {\it one eddy  approximation} to the full energy spectrum of the turbulence.   The model is  local and homogeneous   and does not account explicitly for compressibility and anisotropy; these could however be addressed approximately.  Technically, it consists of a set of coupled ordinary differential equations which are easily solved numerically. 

The turbulence energy spectrum is determined by the rate of energy input from the {\it source} (buoyancy) into
the turbulence. This rate in turn depends   both on the {\it source} parameters and on the {\it turbulence} itself. The physical basis of the model is probably the reason for its success  in predicting observational results, as mentioned above.   

The model   complements numerical simulations by providing {\it   physical understanding} and an easy tool to test for the imprints of the various parameters of the problem. It can be also used a sub-grid model for Large Eddy Simulations (LES). In the context of 
  the  present work we note that
\begin{itemize}
\item[1.] Since CGM96 includes the full turbulence spectrum, the  
turbulent viscosity is expected be larger than the MLT value.
\item[2.] Since  CGM96  has definite prescriptions for the turbulent viscosity and
for the convection timescales -   the ambiguity of the MLT prescription could be avoided.
\end{itemize}

In what follows we report on the {\it results} of this investigation. A more detailed account
will be presented in Goldman (2008).

 \section{The  Effective Turbulent Viscosity in the Presence of Time-Varying Shear} 

Let us consider stationary turbulence that interacts with an external shear. We focus on the case when the rate of energy   input from the shear to the turbulence is small compared to the energy rate fed into the   turbulence by its source (buoyancy in the present case) and does not modify the turbulence.

The turbulence is defined by $F(k)$,    the turbulent velocity spectrum 
\begin{equation} \label{1time  }
   < \vec{v}(t) \cdot \vec{v}(t))> =\int_{k_0}^\infty  F(k)dk 
 \end{equation}
which is related to turbulent energy (per unit mass) spectrum, $E(k)$, by   $F(k)=2 E(k)$.

The wavenumber $k_0$ corresponds to the largest scale in the turbulence energy spectrum with $F(k_0)=0$. The spectral function increases with $k$ and then decreases and approaches a Kolmogorov spectrum, $F(k) \propto k^{-5/3}$. For large enough k, microscopic dissipation truncates the spectrum.

The quantity that controls the interaction between the turbulence and the shear is the turbulent diffusivity $D_T$. It   relates  
 the turbulent stress tensor (caused by the external shear) with the shear tensor.
\begin{equation} \label{def}
 <v_i(t)v_j(t)>= D_T  S_{ij}(t)  
\end{equation}
 The angular bracket denotes ensemble average; the turbulence is stationary, hence the result is independent of   t. The turbulent diffusivity $D_T$, controls also turbulent transport processes.
 
 The turbulent diffusivity $D_T$ has the same dimensions as the turbulent eddy viscosity $\nu_T$. However, the latter controls the non-linear interactions of the eddies, while the former controls the interactions with an external shear.  
 Given a specific turbulence model, these two quantities are defined and the relation between them could be established. In a phenomenological approach, like the MLT, the two 
 are considered to be the same and taken equal (up to some dimensionless coefficient) to the ratio of the eddy size to the turbulent velocity on this scale. In the CGM96 model these are  distinct concepts.
  
For the problem at hand Eq.\,(\ref{def}) takes the form
\begin{equation} \label{rphi}
 <v_r(t)v_{\phi}(t)>= D_T  S_{r\phi} (t)
\end{equation}

Using the   shearing coordinates approach  (Rogallo 1981)
we obtain
\begin{equation} \label{rr}
 <v_r(t)v_{\phi}(t)>= \int_{- \infty}^t <v_r(t)v_r(t')> S_{r\phi} (t') dt' 
 \end{equation}

 The 2-times   ensemble average is related to the turbulence energy spectrum by 
(Monin \& Iaglom 1975;\ Voelk et al. 1980)

\begin{equation} \label{2times3d }
 <\vec{v }(t) \cdot \vec{v }(t'))>=    \int_{k_0}^\infty   F(k) e^{-  \frac{|t-t'|}{\tau_c(k)}  } dk 
 \end{equation}
For the quantity of interest in Eq.\,(\ref{rr}) we get

\begin{equation} \label{2times }
 < v_r(t)  v_r (t')>=    \int_{k_0}^\infty \alpha(k) F(k) e^{-  \frac{|t-t'|}{\tau_c(k)}  } dk 
 \end{equation}
In the expression above, $\tau_c(k)$ denotes   the eddy decorrelation timescale 
and $0< \alpha(k)<1$ , the anisotropy factor of the eddies  defined by

\begin{equation} \label{alpha }
  <v_r(t)v_r(t )>=\int_{k_0}^\infty \alpha(k) F(k)dk .   
 \end{equation}
  
  For strictly isotropic turbulence   $\alpha(k)= \frac{1}{3}$.

For a pure sinusoidal shear
$S(t)=S e^{i\omega t}$ we get an expression for the diffusivity:
 
 \begin{equation} \label{DTomega }
  D_T(\omega)= \int_{k_0}^\infty \alpha(k) F(k) e^{- \omega   \tau_c (k)  }\frac{\tau_c(k)}{1+\left[\omega   \tau_c (k)\right]^2} dk   
  \end{equation}
For slow varying shear ($ \omega   \tau_c (k)<<1$,   for all k),  the
turbulent diffusivity contributed by the full turbulent spectrum is
\begin{equation} \label{DT }
  D_T(0)= \int_{k_0}^\infty \alpha(k) F(k)   \tau_c (k)    dk    
  \end{equation}
  
When the shear is varying faster than the decorrelation timescale of the largest eddies, $\tau_0$, the diffusivity is reduced and  
  can be expressed  as 
  
\begin{equation} \label{eta }
 D_T(\omega)=\eta(\omega   \tau_0)D_T(0)
\end{equation}
where   $\eta(\omega   \tau_0)\leq 1$ is the efficiency factor.

 \subsection{Analytic Approximation of the Efficiency Factor}

The efficiency factor
can be evaluated once the turbulence spectrum and the eddy decorrelation time scale are known. 

If   the turbulence spectrum of CGM96 is approximated by a  Kolmogorov spectrum (even for the largest eddies), it is possible to obtain an {\it analytic  approximation}: 

\begin{equation} \label{approxeta }
 \hskip -0.85 truecm \eta(x)=\frac{20}{ 11 x^2}\left(\frac{3}{4} ln(1+x^2) -\frac{1}{3} +   \frac{1}{x^2} ( 1- \frac{tg^{-1}(x)}{x})\right)
\end{equation}

\begin{figure}
\includegraphics[scale=0.8]{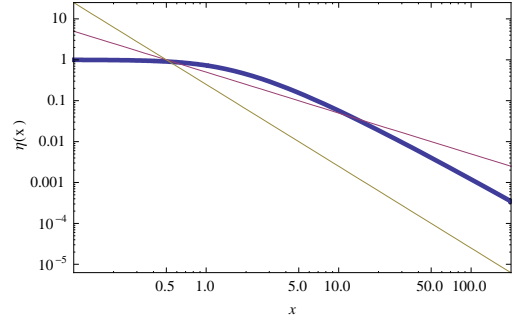}

\caption{{\it Solidline}: the efficiency factor $\eta(x)$. {\it Thinlines}:  $x^{-1}, x^{-2}$, representing linear and quadratic reductions.}thatt
\label{etafig}
\end{figure}

 The efficiency function is shown in figure \ref{etafig} together with efficiencies
that decrease as $x^{-1}$  , and $x^{-2}$, respectively, normalized to unity at x=0.5. As seen, the  decrease  with  x is   quite moderate.   The qualitative shape is similar to that in Goodman and Oh (1997). 

 \section{Application to the Solar Model} 
 
 We have applied the approach outlined above to a binary consisting of two solar mass stars.
and used the turbulence model of CGM96 to compute the relevant quantities. The details
 will be published elsewhere (Goldman 2008). The main results are presented below. 

\subsection{Tidal Timescales for a Stationary Shear} 

The basic tidal timescale  $T_c$ the circularization time,  is inversely proportional
to a weighted   average, over the convective zone, of the product  $\rho D_T$:
 
\begin{equation} \label{tc}
<\rho D_T>= \int_{0.7} ^1 \rho D_T  x^8 dx
\end{equation}  
with $x= r/R_{\odot}$.

For a stationary shear

\begin{equation} \label{tcs}
<\rho D_T(0)>= \int_{0.7} ^1 \rho D_T(0)  x^8 dx
\end{equation}  

The numerical solution of the solar model of CGM96, yields  
\begin{equation} \label{tcsval}
<\rho D_T(0)>= 2.4   \times 10^{11}  \ gr cm ^2 s ^{-1}
\end{equation}

  This is
is a factor of $\sim 4.5$ larger than the corresponding value computed within the MLT framework. For a binary  consisting of solar mass members, with orbital period of $15 $ days
 the circularization timescale is  
  $T_c(0)= 1.7 \times 10^{11}$ yr. So, the discrepancy with the observational circularization timescale is alleviated by the same factor and is now  "only"   a factor of $\sim 12$ too large. 
  
  The contribution to  $<\rho D_T(0)>$ peaks in the middle of the convective zone, reflecting the facts that the  density is a decreasing function of the radius and the diffusivity  $D_T(0)$ is an increasing function.

The increase of $<\rho D_T(0)>$ in comparison to the MLT, is not as large as might have been  anticipated. According to CGM96, taking into account the full turbulent spectrum enhances   
the convective flux for a given super adiabatic gradient, by an order of magnitude, compared to the MLT. Since the value of
 the
convective flux is imposed by the radiative core, the super adiabatic gradient is smaller, than in the MLT models. This
compensates partially, the increase  stemming from the inclusions of the full turbulence spectrum.

\subsection{Tidal Timescales for a Dynamic Shear} 

We have  computed at each radius the local
 eddy decorrelation time scale $\tau_0$, and from it the local
 efficiency factor which was then used to evaluate  $<\rho D_T>$ from Eq. \ref{tc}.
 
 We obtained  that for a 15 day binary, the weighted efficiency is $\simeq 0.4$. This {\it very mild} reduction leads  to a tidal time scale which 
 is too long by a factor   of $\simeq 30$, compared to a factor of $\simeq 1000$, according to the MLT.

\section{Discussion} 

The results indicate that the use of a physically based model for turbulence alleviates considerably the discrepancy between the observational and predicted tidal time scales. The 
approximation, used here, of  a Kolmogorov spectrum even for the largest eddies, underestimates the
diffusivity. Therefore,
we intend to perform a more rigorous, non-analytical computation. 

\acknowledgements 
The author thanks Tsevi Mazeh for interesting discussions and the  Institute of Astronomy and Astrophysics, Tel Aviv University for a research grant.

\end{document}